\documentclass[twocolumn,prb,showpacs,superscriptaddress,amsmath,amssymb,natbib]{revtex4}

\usepackage[dvips]{epsfig}
\usepackage{graphicx}
\usepackage{dcolumn}
\usepackage{bm}
\usepackage[english,german]{babel}
\usepackage{amsmath}
\usepackage{amssymb}
\usepackage[latin1]{inputenc}

\newcommand{\mat}[1]{\boldsymbol{#1}}

\begin{document}

\title{Computing absolute free energies of disordered structures by molecular simulation}

\author{T. Schilling}
\email{schillit@uni-mainz.de}
\affiliation{\selectlanguage{german}
  Institut f"ur Physik, Johannes Gutenberg-Universit"at, Staudinger Weg 7, D-55099
  Mainz, Germany
}

\author{F. Schmid}
\email{friederike.schmid@uni-mainz.de}
\affiliation{\selectlanguage{german}
  Institut f"ur Physik, Johannes Gutenberg-Universit"at, Staudinger Weg 7, D-55099
  Mainz, Germany
}

\selectlanguage{english}

\date{\today}

\begin{abstract}
We present a Monte Carlo simulation technique by which the free energy 
of disordered systems can be computed directly. It is based on 
thermodynamic integration. The central idea is to construct
an analytically solvable reference system from a configuration which is 
representative for the state of interest. The method can be applied to 
lattice models (e.g., the Ising model) as well as off-lattice molecular
models. We focus mainly on the more challenging off-lattice case. 
We propose a Monte Carlo algorithm, by which the thermodynamic 
integration path can be sampled efficiently. At the examples of the 
hard sphere liquid and a hard disk solid with a defect we discuss 
several properties of the approach.  
\end{abstract}


\maketitle

The fundamental equation $S=f(U,V,\{N_\alpha\})$, which connects the entropy 
$S$ with the internal energy $U$, the volume $V$, and the numbers $N_\alpha$ of
particles of type $\alpha$, contains all information about a system that is 
accessible within classical thermodynamics. Other thermodynamic potentials
such as e.g.~the free energy are related to the fundamental equation by
Legendre transform, and hence they equally contain this 
information \cite{jelitto89}. Therefore there is large interest in 
computing free energies in many areas of science, {\em i.e.}, statistical 
physics, materials science, theoretical chemistry, and 
biology \cite{chipot07}.

There are only very few, special cases in which the free energy of a system 
can be computed directly: Either the accessible phase space volume can be 
enumerated completely (as e.g.~for a lattice gas model on a small lattice), 
or the problem can be solved analytically in the first place (as e.g.~for the 
ideal gas). In all other cases one must resort to approximations or to 
computer simulations. Unfortunately, the latter only give access to free 
energy derivatives and free energy differences. Several advanced techniques 
have been developed that allow to relate free energies of different state 
points to each other, and a large body of literature has been written on 
this topic \cite{chipot07,earl05,okamoto04,Panagiatopoulos00,dePablo99,
Wilding00a,Wilding00}. 
Nevertheless, comparing the free energies of arbitrary systems remains 
a challenge, and alternative approaches that allow to determine 
the absolute free energy for each individual system are clearly of 
interest. 

On principle, absolute free energies can be obtained by connecting 
the system of interest with a reference system of known free energy. 
In this letter, we propose a general strategy for the construction of 
analytically solvable reference systems, that can be connected with
a wide class of structures {\em via} thermodynamic integration.

Thermodynamic integration \cite{frenkelsmit02,LandauBinder} is a widely
applied method to determine free energy differences.
The basic idea is the following: Consider a system of $N$ particles with 
a Hamiltonian $H(\mat{r}^N, \mat{p}^N, \varepsilon)$, which explicitly 
depends on some parameter $\varepsilon$.  In order to obtain an
expression for the free energy of the system, one uses the relation  
${\partial F}/{\partial \varepsilon} = 
  \left\langle {\partial H(\varepsilon)}/{\partial \varepsilon}
\right\rangle$,
where $\langle \ldots \rangle$ denotes the thermodynamic average.  
Here and in the following, we set $k_B T = 1$.
In general, $\langle \partial H (\varepsilon)/\partial \varepsilon \rangle$,
is directly accessible in a simulation. Thus the expression above
can be used to evaluate the free energy difference between two systems
at different $\varepsilon$: One samples
$\left\langle {\partial H(\varepsilon)}/{\partial \varepsilon}\right\rangle$
for a range of $\varepsilon$ and integrates
\begin{equation}\label{eq:TI1}
    \Delta F = F(\varepsilon_1)-F(\varepsilon_0) 
  = \int_{\varepsilon_0}^{\varepsilon_1} \: \mathrm{d}\varepsilon^\prime\,
      \bigg\langle
      \frac{\partial H(\varepsilon^\prime)}{\partial\varepsilon^\prime}
    \bigg\rangle_{\varepsilon^\prime}\quad .
\end{equation}
If the free energy is known for one $\varepsilon_0$ (reference system),
the method can be used to calculate absolute free energies for a whole
range of $\varepsilon$. However, it is crucial
that the evolution of $\langle \partial H (\epsilon)/\partial \varepsilon \rangle$ 
on the integration path is reversible, {\em i.e.}, no phase transition of 
first order may be crossed.  This limits the choice of suitable integration paths 
and reference systems.  For gases the ideal gas is a useful reference system, for 
crystals the ``Einstein crystal'' (a crystal where the particles are bound to sites 
of a fixed lattice by harmonic springs \cite{einstein07,frenkelladd84}). To the best
of our knowledge, no general reference system has been introduced so far that can
be used for arbitrary dense disordered systems.

Our central idea to remedy this situation is very simple. We propose to take a 
configuration that is representative for the structure of interest (obtained 
{\em e.g.} within a typical simulation of an equilibrated system) and to construct 
a reference system by first 'pinning' this configuration with suitable external 
fields, and then switching off the internal interactions. In the remainder of 
this letter, we will show how this idea can be exploited to evaluate absolute 
free energies in practice.

For the purpose of illustration, we begin by considering the Ising model
$H_0 = -J \sum_{\langle ij \rangle} s_i s_j$,
where $\langle ij \rangle$ denotes neighbouring $i$ and $j$ and 
$s_i = \pm 1$.  To evaluate the free energy $F_0$ at a given temperature, 
we simulate the system until it is equilibrated, and then pick one 
typical configuration $\{s_i^R\}$ as 'representative' 
reference configuration. The reference system is then defined by
the Hamiltonian
\begin{equation}
\label{ref:ising_reference}
  H_{\rm ref}(\varepsilon) =  - \varepsilon \sum_i s_i s_i^R \quad ,
\end{equation}
and its free energy can be computed easily, $F_{\rm ref}(\varepsilon) = 
-N\ln\left(2 \cosh \left(\varepsilon\right)\right)$. To establish the
connection with the original system, we procede in two steps: First 
we define an intermediate model
$H'(\varepsilon) = H_0 + H_{\rm ref}(\varepsilon)$, which reduces to
$H_0$ at $\varepsilon = 0$. The free energy difference 
$\Delta F_1(\varepsilon) = F_0 - F'(\varepsilon)$ between the
original system and the intermediate system can be calculated 
for arbitrary $\varepsilon$ by thermodynamic integration, using
$\langle \partial H'/\partial \varepsilon \rangle 
= - \langle \sum_i s_i s_i^R \rangle$.  We choose $\varepsilon$ large
enough that the spins in the system $H'(\varepsilon)$ hardly
fluctuate about the reference value. The second step is to connect the
intermediate system to the reference system. The free energy difference 
between the two systems at the same value of $\varepsilon$, 
$\Delta F_2(\varepsilon) = F'(\varepsilon) - F_{\rm ref}(\varepsilon)$,
is evaluated by carrying out a simulation with additional 
Monte Carlo (MC) moves that switch on and off the interaction $J$ 
according to a Metropolis criterion. We obtain
$\Delta F_2(\varepsilon) = - \ln (P_{\rm on}/P_{\rm off})$,
where $P_{\rm on,off}$ is the fraction of configurations 
with interactions switched on (rsp. off) in the simulation.
Combining everything, we can finally calculate the absolute
free energy of the target system $H_0$,
$F_0 = F_{\rm ref}(\varepsilon) + \Delta F_1(\varepsilon)
+ \Delta F_2(\varepsilon)$.

Now we transfer this idea to off-lattice particle models. For 
clarity, we only discuss monatomic liquids and solids in the 
$NVT$ ensemble in the following. Our method can easily 
be generalized to molecular systems, and, as we shall demonstrate 
below, to constant pressure simulations. Furthermore, we 
disregard the kinetic contribution to the free energy, 
which can be evaluated trivially \cite{HansenMcDonald}.

Let configurations be characterized by a set of coordinates
$\{\mat{r}_i\}$ and the configurational energy be given by
a Hamiltonian $H_0 = U(\{\mat{r}_i \})$. 
To calculate the free energy of a given, arbitrary equilibrium 
structure, we choose a 'representative' configuration 
$\{\mat{r}^R_i\}$, obtained, {\em e.g.}, from a simulation 
of an equilibrated system, and construct a reference system 
by imposing local potentials
\begin{equation}\label{RefHamilton}
H_{\rm ref} (\varepsilon) = \varepsilon\sum_i 
\Phi\left(\frac{|\mat{r}_i - \mat{r}^R_i|}{r_{\rm cutoff}}\right) 
\quad , 
\end{equation}
that pin the particles' positions $\mat{r}_i$ to the reference 
positions $\mat{r}^R_i$. Here $\Phi$ defines attractive 
potential-wells centered at each position $\mat{r}^R_i$, with
$\Phi(x) <0$ for $x < 1$ and $\Phi \equiv 0$ elsewhere.
Note that particle $i$ can only be trapped by well $i$ and 
not by the other wells. To make the 
particles indistinguishable, as they should be, we allow them 
to swap identities ({\em i.e.}, labels $i,j$) at regular 
intervals during the simulations.  We will show below that 
such identity swaps are also necessary to equilibrate the 
system efficiently. 

The (configurational) reference free energy is given by
\begin{equation}
  \frac{F_{\rm ref}(\varepsilon)}{N} 
    = \ln\left(\frac{N}{V}\right) 
      - \ln\left(1 +\frac{V_0}{V} g_{\Phi}(\varepsilon)\right)\quad ,
\end{equation}
where $V_0$ is the volume of the sphere of radius $r_{\rm cutoff}$ and 
$
g_{\Phi}(\varepsilon) := {\rm d} \int_0^1 dx\;x^{{\rm d}
  -1}\left(e^{-\varepsilon\Phi(x)}-1\right) 
$
for a ${\rm d}$-dimensional problem. In our simulations, we mostly
used a linear well potential, $\Phi^{\rm linear}(x) = x-1$.
In this case one has
$ g_{\Phi}(\varepsilon) = 
{\rm d}/{\varepsilon^{\rm d}}\left(e^{\varepsilon}  \:
- \sum_{k=0}^{\rm d}{\varepsilon^k}/{k!}\right)$.
As before, we also define an intermediate model 
$ H'(\varepsilon) = H_0 + H_{\rm ref}(\varepsilon)$,
and evaluate the free energy difference between the reference 
system and the intermediate system at high $\varepsilon$
with a MC simulation where the interaction $H_0$ is
switched on and off (if necessary, in several steps).
The free energy difference between the target system 
and the intermediate system is computed by sampling
$\partial F'/\partial \epsilon = 
\langle \sum_i \Phi(|\mat{r}_i - \mat{r}^R_i|/r_{\rm cutoff}) \rangle$
and performing a thermodynamic integration. The remaining
challenge is to devise an algorithm for sampling
the intermediate model efficiently for
arbitrary $\varepsilon$.

Before describing such an algorithm, we briefly comment
on the relation between our method and the Einstein crystal
method to determine free energies of solids \cite{frenkelladd84}.
In the Einstein crystal method, the particles are not 
swapped, and the reference system is a regular lattice 
of harmonic wells with infinite cutoff radius. This works
well as long as the particles in the target system 
stay close to their respective well positions. In a liquid,
where their mean-square displacement diverges, 
$\langle \sum_i \Phi(|\mat{r}_i - \mat{r}^R_i|/r_{\rm cutoff}) \rangle$
diverges as well for small $\varepsilon$ and can no 
longer be sampled. Therefore the introduction of a 
finite cutoff is crucial. We note that our method can also 
be used to evaluate free energies of crystals. 

\begin{figure}[t]
\begin{center}
\includegraphics[width = .7 \columnwidth] {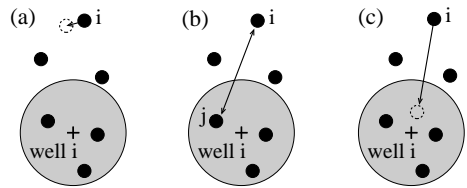} 
\vspace*{-0.1cm}
\caption{Sketch of moves in our Monte Carlo algorithm.
(a) Simple particle displacements. (Could be replaced
{\em e.g.}, by short Molecular Dynamics runs.)
(b) Smart particle swaps.
(c) Smart particle relocations.
See text for explanation} \label{fig:moves}
\end{center}
\vspace*{-0.2cm}
\end{figure}

Setting a finite range for the reference potential, however, 
introduces a different problem: The particles need to find 
their respective wells of attraction. We therefore introduce 
two MC moves that help particles $i$ explore their 
well $i$ (Fig.~\ref{fig:moves}).
One move (Fig.~\ref{fig:moves} b) swaps particles in
a smart way. It works as follows:

\noindent {$\bullet$}
\parbox[t]{0.95 \linewidth}{
  Pick a random particle $i$ and find the set of particles $\{n_i\}$
  that are within the attraction range of well $i$.
}
\noindent {$\bullet$}
\parbox[t]{0.95 \linewidth}{
  If particle $i \notin \{n_i\}$: pick a particle $j$ from $\{n_i\}$,
  and swap $i$ and $j$ with the probability 
    $\min\{1,\frac{n_i}{N}e^{-\Delta H'}\}.$
}
\noindent {$\bullet$}
\parbox[t]{0.95 \linewidth}{
  Otherwise: pick a particle $j$ from all particles  \\
  - if $j \notin \{n_i\}$: swap with probability
      $\min\{1,\frac{N}{n_i}e^{-\Delta H'}\}$. \\
  - if $j \in \{n_i\}$: swap with probability
      $\min\{1, e^{-\Delta H'}\}.$
}
Here $\Delta H'$ is the difference of the energies
(according to the intermediate model) of the old and 
new configuration. This algorithm promotes particle swaps 
that bring particles close to their respective well and 
nevertheless satisfies detailed balance.  

\begin{figure}[t]
\begin{center}
\includegraphics[width = .8 \columnwidth] {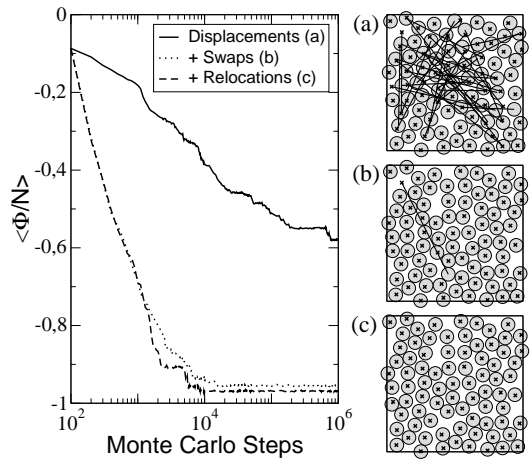} 
\vspace*{-0.1cm}
\caption{Illustration of the effect of the
moves of Fig.~\protect\ref{fig:moves} on the
equilibration of a system of 80 hard disks 
(diameter $D$) at a density $\rho = 0.8/D^2$, 
after switching on linear well potentials with 
strength $\varepsilon = 50$ ($r_{\rm cutoff} = 2 D$).
Swap moves and relocation moves (one per bead) were 
attempted one per 100 MC sweeps.
Left: Evolution of $\langle \Phi \rangle$ 
in simulations that include 
different moves as indicated. 
Right: Corresponding final configurations.
Circles indicate particle positions, crosses
give well positions. Particles and their respective
wells are connected by straight lines.
} 
\label{fig:equil}
\end{center}
\vspace*{-0.2cm}
\end{figure}

The other move (Fig.~\ref{fig:moves} c) relocates particles 
$i$ with a bias towards the neighborhood of their well $i$:

\noindent {$\bullet$}
\parbox[t]{0.95 \linewidth}{
Pick a random particle $i$ (with position $\mat{r}_i$).
}
\noindent {$\bullet$}
\parbox[t]{0.95 \linewidth}{
  Choose a new position $\mat{r}'_i$ from a given 
  (biased) distribution $P_i(\mat{r'}_i) 
   = \exp(-W(|\mat{r'}_i - \mat{r}_i^R|))$. 
}
\noindent {$\bullet$}
\parbox[t]{0.95 \linewidth}{
 Relocate the particle from $\mat{r}$ to $\mat{r}'_i$ with probability
  $\min\{1,P(\mat{r}_i)/P(\mat{r}'_i) \: e^{-\Delta H'}\}$.
}
Obvious choices for $W(r)$ which we have tested are
$W(r) = \varepsilon \Phi(r/r_{\rm cutoff})$,
or $W(r) = {\rm const.}$ for $r < r_{\rm cutoff}$. 
At high $\varepsilon$, the relocation move helps to 
overcome trapped situations where most particles are bound 
to a well, and a few cannot escape from a local cage.
To illustrate the effect of
the different moves, Fig.~\ref{fig:equil} shows the evolution 
of the observable $\langle \Phi_i \rangle$, averaged over
all particles $i$, in a two dimensional system of
hard disks, after $\varepsilon$ had been raised 
from zero to a high value. In a MC simulation that includes 
only random particle displacements, the system is far from 
equilibration after one million MC sweeps (a). 
The smart swap moves speed up equilibration considerably, 
but the system gets trapped in a configuration where one 
particle cannot enter its well (b). This problem is 
solved by including smart relocation moves (c).

\begin{table}[ht]
  \caption{\label{tableRes}Results for the free
    energy of hard spheres. ${F/N}_{\rm CS}$ is the 
    value according to the Carnahan-Starling equation of state \cite{carnahan69}. 
    a) linear potential $\Phi$, liquid reference state. 
    b) linear $\Phi$, hcp reference state. c) harmonic $\Phi$, liquid
    reference state.}
  \begin{ruledtabular}
    \begin{tabular}{ccc}
      $N/V$ & $F/N$ & $(F/N)_{\rm CS}$ \\
      \hline
      $0.25$ & $0.62(0)$ & $0.625$ \\
      $0.5^{a)}$ & $1.54(1)$ &$1.544$ \\
      $0.5^{b)}$ & $1.54(0)$ &$1.544$\\
      $0.5^{c)}$ & $1.54(9)$ &$1.544$\\
      $0.75$     & $3.00(9)$ &$3.005$ \\
    \end{tabular}
  \end{ruledtabular}
\end{table}
\vspace*{-0.2cm}

We will now demonstrate the power of our approach at
a few examples. They are not meant to be self-sufficient
scientific studies -- the simulations were carried out
on simple workstations with poorly optimized test programs
in relatively short time (less than a week). We are 
aware that a careful analysis of the finite-size effects 
should be done in all cases. Here we only intend to
illustrate some properties of the algorithm.

We have studied hard spheres in two (2d) and three dimensions (3d). 
For the remainder of
this letter we use the particle diameter $D$ as unit of length.
Table \ref{tableRes} shows results for the free energy of a 
3d liquid of hard spheres. The simulations were performed on 
a system of $N=256$ particles, using 50 values of $\varepsilon$ and 
$6 \cdot 10^5$ MC sweeps for each value at $N/V=0.25$ and $N/V=0.5$, and 200 values of 
$\varepsilon$ times 1 Mio. MC sweeps at $N/V=0.75$. 
The results agree with
the values obtained by integration of the Carnahan-Starling equation of 
state \cite{carnahan69} within the error bars. For $N/V = 0.5$ we compared the cases
a) linear potential $\Phi$ and liquid reference state,
b) linear $\Phi$ and crystalline reference state, and c) harmonic $\Phi$
and liquid reference state. Within the error-bars these variations 
produce the same result. However, for more accurate calculations the 
linear potential seems to be most useful, because the particles get 
trapped most efficiently. In case b) we did not see a
hysteresis on increasing/decreasing $\varepsilon$. Apparently, the trapping
of particles in a crystalline array of wells is not associated with
a phase transition at this density. This will presumably be different
closer to liquid/solid coexistence. Nevertheless we can conclude that
our method is quite robust and may work even if the reference configuration 
is not 'ideal', {\em i.e.}, not representative of the target structure.

Next we show an example for the application of the method to 
dense disordered systems, where the dynamics is driven by 
cooperative processes. We studied hard disks in 2d 
up to densities where the equilibrium phase is a solid, and enforced a 
vacancy defect by taking one particle out of an otherwise 
ordered configuration. These simulations 
were carried out at constant pressure $P$ in a rectangular simulation box 
of varying area, but fixed side ratio $1:\sqrt{3/4}$, to accomodate a 
triangular lattice. The defect is then stable, but highly mobile.

\begin{figure}[t]
\begin{center}
\includegraphics[width = .8 \columnwidth] {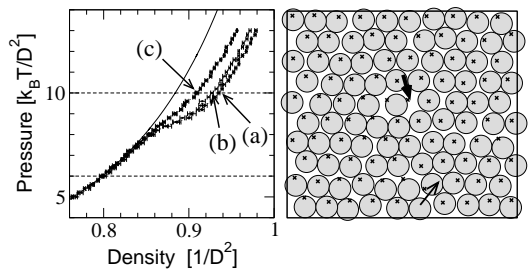} 
\vspace*{-0.1cm}
\caption{
Characterization of the dense two-dimensional systems 
discussed in the text. Left: Pressure vs. density
as obtained from constant pressure simulations at $\varepsilon=0$. 
(a) $N=100$ particles, expanded from an ordered solid phase,
(b) $N=99$ particles, expanded from an ordered solid phase with one vacancy,
(c) $N=100$ particles (diameter $D$), compressed from the fluid phase. 
The solid line shows the theoretical
estimate $P = \hat{\rho}/(1-\frac{\pi}{4} \hat{\rho})^2$ with
$\hat{\rho} = (N+1)/\langle V \rangle$. 
Right: A configuration with one vacancy at the
beginning (crosses) and the end (circles) of a MC 
run. The thin and thick arrows mark the position
of the defect at the beginning and the end.
} 
\label{fig:defect}
\end{center}
\vspace*{-0.2cm}
\end{figure}

We compare three different structures (Fig.~\ref{fig:defect}): 
An ordered solid (a), an ordered solid with a vacancy (b),
and a metastable disordered jammed phase (c), which was obtained
by compressing the system from the fluid phase. Free energy 
calculations were carried out at $P=10$ for these
three cases, and additionally at $P=6$, in the fluid regime.
To calculate the free energy in the enthalpic ensemble, we use a
reference system that is defined in terms of scalable coordinates 
({\em i.e.}, the positions of the well centers are rescaled along
with the particle coordinates if the volume of the system changes), 
and pin the volume of the system by an additional term 
$\varepsilon (V - V_{\rm ref})^2$ in the reference Hamiltonian.
The thermodynamic integration was carried out for 
$\varepsilon = 0\ldots 40$ using 66 values 
of $\varepsilon$, and 10 Mio.~MC sweeps for 
each $\varepsilon$. The resulting free enthalpies 
$G$ can be related to the chemical potential $\mu$ by virtue 
of the thermodynamic relation $G = \mu N$.

To set the frame, we show in Fig.~\ref{fig:defect} a) the pressure-density
curves for the cases a),b), and c). In the fluid regime 
($\rho \le 0.8$), they can be fitted nicely with the theoretical 
estimate \cite{helfand60} $P = \hat{\rho}/(1-\frac{\pi}{4} \hat{\rho})^2$ with
$\hat{\rho} = (N+1)/\langle V \rangle$. Fig.~\ref{fig:defect} b) 
illustrates the mobility of the defect at pressure $P=10$.
It should be noted that in the solid regime, the center of mass motion 
of the complete system is not sampled well, because individual
particles stay close to their lattice sites. A similar problem is 
encountered in the Einstein crystal method and has 
lead to the development of the `Einstein molecule'
\cite{veganoya07}, where the reference crystal is defined in terms 
of relative coordinates. This idea can easily be transfered to our 
approach. Here, we ignore the center-of-mass correction, 
because it is smaller than our statistical error.

At $P=6$, the free energy calculation yields the free enthalpy
per particle $\mu = 8.997 (2)$, which is reasonably close to 
the theoretical estimate $\mu = 9.047$ obtained by integrating 
the theoretical equation of state. At $P=10$, we found
$\mu_{\rm solid} = 13.617(2)$ in the solid state, and
$\mu_{\rm jam} = 13.675(2)$ in the jammed state, which establishes that the 
solid is indeed the stable phase. For the system with one
defect, we obtained the total enthalpy $G_{\rm defect} = 1361.7(2)$. 
This result can be used to estimate the core free energy of the vacancy
$\mu_c = G_{\rm defect}- \mu_{\rm solid} N  + \ln (N)= 7.1(3)$, 
which corresponds to a relative vacancy frequency of roughly $10^{-3}$.
(For comparison, the frequency of vacancies at liquid/solid 
coexistence in 3d is roughly $10^{-4}$ \cite{pronk01}.)
Probably $\mu_c$ is largely overestimated 
due to finite size effects, hence the value given above is at 
best an upper bound. More detailed studies shall be carried
out in the future. Here, the example mainly serves to illustrate
the use of our approach in situations where free energies
are difficult to access with other methods.

In summary, we have introduced a general method to compute 
absolute free energies for a wide range of structures.
We have illustrated the method for monatomic simple systems, 
but we believe that it can be applied equally well to molecular 
fluids and mixtures.  We anticipate that our method will be useful 
to calculate free energies of systems that are not directly 
connected with the ideal gas, such as liquid crystal phases, 
membranes, or proteins in solution. Also defects in solids 
seem to be a promising field of application. From a fundamental 
point of view, it should be interesting to study how well the
method can be applied to glassy systems, which
have not just one, but a whole set of 'representative' 
configurations, one for each local minimum in the 
rugged free energy landscape.

\vspace{-0.8cm}

\begin{acknowledgments}

\vspace{-0.2cm}
We thank M. Oettel for stimulating discussions.
We are grateful to the DFG (Emmy Noether Program) for financial
support. 
\end{acknowledgments}

\end{document}